# THE ALIGNMENT OF PUBLIC RESEARCH SUPPLY AND INDUSTRY DEMAND FOR EFFECTIVE TECHNOLOGY TRANSFER: THE CASE OF ITALY[1]

*Giovanni Abramo* [a,b,*] *and Andrea D'Angelo* [b]

**Abstract**

Italy lags quite behind vis-à-vis other industrialized countries, in public to private technology transfer. One of the possible causes might be the mismatch between new knowledge supplied by public research and industry demand. We test this hypothesis through a survey of leading public research scientists in four high-tech sectors. The findings show that most research project results seem to be of immediate industrial interest, which contrasts with the low patent and licensing performances of Italian public research institutions. For one third of all the results of the research, there are no Italian companies able to exploit them. The same, however, is not true for the remaining results, which shows that the misalignment between public supply and industry demand alone cannot account for poor technology transfer. What emerges from our investigation is that a closer coordination of research policy and industrial policy is required, as well as closer attention to initiatives which may support the transfer of public research results to domestic industry.

[a] Consiglio Nazionale delle Ricerche (National Research Council)
[b] Laboratorio di Studi sulla Ricerca e il Trasferimento Tecnologico (Laboratory for Studies of Research and Technology Transfer) at "Tor Vergata" University
* Corresponding author: Università degli Studi di Roma "Tor Vergata", Dipartimento di Ingegneria dell'Impresa, Via del Politecnico 1 – 00133 Rome, Italy; tel +39 06 72597362 fax +39 06 72597305; abramo@disp.uniroma2.it

# 1. Introduction

There is an ever increasing interest in Italy regarding the transformation process of the role of universities and public research institutions in the country's economic system, and their contribution to industrial competitiveness, at both local and national levels.

In what is known as the *knowledge economy*, science has become more crucial to innovation processes than it ever used to be, especially in high-tech sectors. As a result, universities and public research institutions have assumed an increasingly decisive role in industrial competitiveness, economic growth and employment. This has led not only to increased discussion within the international scientific community but also to governance and policy actions by public research institutions.

According to Tassey, 1991, the accelerated growth of some industrialized nations is partly the result of the complementary role effectively played by governments, which have provided domestic economies with an adequate technology base and have enabled its broad and rapid utilization. There is a widespread awareness that an increase in the scale and scope of research and an improvement in the productivity of research organizations alone cannot maximize the socio-economic returns on R&D spending. Technology transfer, particularly from the public sector to the private sector, has become increasingly central in the research policy agenda of a growing number of countries. The topic has also generated great interest among academic researchers. Organizations such as the Association of University Technology Managers (Autm) in the U.S., the University Companies Association (UNICO) and the Association for University Research Industry Links (AURIL) in the U.K. have helped to promote technology transfer activities by publishing benchmarking surveys, which have been used by researchers to explore key research questions relating to the drivers of effective public-to-private technology transfer. In the last 25 years, the terms "technology transfer" and "technology diffusion" have appeared in the titles of hundreds of articles and books. A large part of such extensive multidisciplinary literature has been reviewed and reduced to manageable proportions by Bozeman (2000); interested readers can refer to his work.

There are various mechanisms through which the results of research by public laboratories are disseminated to industry: publications, conferences, patents, license and know-how contracts, spin-offs, joint research, consultancy, staff mobility, training etc. The different modes of codification and transfer of new knowledge contribute, to different degrees, to supporting the competitiveness of productive systems (Cohen et al., 2002a) and thus to the socio-economic development of a country. All else being equal, the more a particular form of codification contributes to making proprietary the newly produced technical knowledge of immediate interest to industry, the more powerful its potential impact on industrial competitiveness. This awareness has certainly led many governments to strengthen their technology transfer policies and many public research institutions to adjust their *purposes* and realign their strategies, organizational structures and management systems. The American and Canadian research systems, followed by the British and Dutch, were pioneers in this process. Italy, instead, as we will show later, lags quite behind in technology transfer intensity.

Three main generic factors may determine Italy's comparatively low public-private technology transfer results. They are: i) low demand from industry; ii) a mismatch between public supply and industry demand; and iii) an ineffective technology transfer system. Probably, the interplay of all three factors is the main cause of the current Italian situation. With regard to the first factor, the difficulty in transferring public patents appears peculiar within the Italian economy, which regularly experiences a negative technology balance of payments, especially with respect to patenting transactions[1]; which proves that the technology demand from Italian industry is higher than the domestic supply. As for the third factor, Abramo, 2007 conducted an in-depth analysis of the Italian technology transfer system and identified its main weaknesses in the inadequate coordination and integration of policies and actions by the three main players, i.e. government, research institutions and industry.

The objective of this paper is to verify the second factor, i.e. whether a mismatch exists between the private demand and the public supply of "new knowledge"[2]. This hypothesis was left unverified by Abramo (2006): in his view, it is reasonable to think that, as the top performance in scientific productivity would seem to suggest, the Italian public research system has not undergone a "high tech de-specialization" similar to that which industry has experienced over the last few years. This may have been favored by the strong ties that national public research systems tend to establish with the public research systems abroad, sometimes stronger than those with domestic industry.

The sectoral specialization of Italian industry (low and middle-low tech) may then give rise to doubts on possible supply-demand mismatch. In order to verify that hypothesis, an empirical survey was carried out, in which the "suppliers", i.e. the public researchers, were asked about the potential exploitability of their current research findings by national industry. The interviews also provided important ancillary information, especially on the ways in which new knowledge is codified, which highlighted interesting characteristics of the current profile of Italian public research. The remaining part of the paper is articulated as follows. The next section presents evidence from the literature of the scarce public to private technology transfer in Italy. The third section shows our research methodology. In the fourth section, we present the most salient findings from the processing of the data, a significance analysis, an analysis of response distributions within the sample, and cross-correlations between the different elements of the study. The final section contains an interpretation of the empirical findings, from a perspective of integration between public research and industrial systems. The usual cautions apply in interpreting findings of empirical surveys of this kind, as personal opinions of interviewees may be biased by various factors.

## 2. Public-to-private technology transfer in Italy

The Italian technology transfer system and its performance need to be interpreted within the historical, cultural and structural characteristics of the context. Differently, from most Oecd countries, R&D spending in the public sector is higher (51%) than in the private sector (in the other G7 countries the average is under 30%). To this anomaly, can be added at least another two of a structural nature, rendering the process of public-private technology transfer, on the one hand, undoubtedly more difficult, and, on the other hand, even more necessary. The first is the progressive demise of the Italian high-tech industry, as shown by the worsening specialization indexes of the Italian economy and the revealing historical analysis of Gallino (2003)[3]. The second is the composition of the Italian industrial system, characterized by a disproportionate ratio of small and micro companies.

The effects deriving from the synergistic interaction of these three anomalies are aggravated by the cultural humus of poor interaction and collaboration amongst public research institutions and industry. The recently sought reversal in this tendency encounters difficulty in taking off due, to the structural distance that has been created between the two systems.

The empirical evidence emerging from the studies effected so far and presented below on the Italian case shows: i) a strong tendency by the Italian public researcher, compared to his/her counterpart in other countries, to codify new knowledge as articles rather than patents; and ii) a low intensity of public to private technology transfer.

The comparison in the number of scientific publications per researcher in the public sector in various countries is complicated by the absence of statistics relating to the distribution of publications by discipline. Adopting the usual cautions in interpreting the following, elaborations of Oecd figures (2006) show[4] that the average number of annual publications of the Italian public researcher, 0.82, is in line with that of their British and American counterparts, respectively 0.86 and 0.81, but distinctly superior to that in France, 0.51; Germany, 0.55; and Japan, 0.32. This means that the average Italian public scientist has a research productivity 62% higher than the French one, 49% higher than the German, and 160% higher than the Japanese[5].

Since it is difficult to ascribe this superiority entirely to the different sectors of scientific investigation[6] or to higher productivity in Italy in research activities[7], it is legitimate to hypothesize, rather, that the Italian public researcher tends to utilize publication as a form of codifying new knowledge far more than his/her foreign counterpart. This hypothesis is borne out by at least two grounds. The first reveals that scientific publications in American universities in the years 1995 – 2000 decreased by 10% against an increase in research costs for the same period by 22%, at constant prices. In the same time frame, in Canada scientific publications decreased by 9%, in the Netherlands by 5%, and in the United Kingdom by 1% (NSB, 2002, Oecd, 2003). Vice versa, the number of patents filed and licensed by American and Canadian universities increased in the same period respectively by 220% and 160% (Autm, 2003).

Italy, on the other hand, registered, in the same period, the highest rate of annual growth in publication amongst the G7 countries. As for patents, in order to have a comparable order of magnitude of patent production by the whole of the Italian universities (more than 70), in 2001[8,] this was more or less equal to that of the University of Wisconsin; the overall figure for universities and public research bodies is inferior to that of the Massachusetts Institute of Technology, MIT, which has a research budget more or less equal to that of the Italian Research Council (CNR) alone (Abramo and Pugini, 2005, Autm, 2004).

The CNR, the Italian research body with the highest patent production in the period 1982 – 2001, detaining 59,1% of the total of European and American patents in the entire public research system, about nine times that of all the Italian universities put together (Piccaluga and Patrono, 2001), in the period 1996 – 2001 registered a contraction of 35% in the number of patents filed per unit of cost in research compared to 1989 – 1995 (Abramo, 2006). The CNR, until 1985, filed more patents than did the MIT; in 1994 the intensity of patents of the average CNR researcher slid to 61% of his/her MIT counterpart (Abramo, 1998) and in 2001 to 26% (Abramo, 2006). Italian universities, despite their scant patent production, presented instead a positive trend from 1997 to 2001, after which there was a reversal, following the introduction of the new law of ownership of patent rights in public research laboratories (Baldini et al.,

2006).

In reality, the recourse to patenting by Italian university research personnel is very much higher than that revealed by the survey of patents owned by the universities. Balconi et al., 2003, demonstrate that the patents filed by university inventors between 1979 and 1999 with the European Patent Office, EPO, were in total 1426, but only 40 were university owned, less than 3% (while 77% were owned by companies). The same phenomenon can be seen for the CNR, where EPO patents by CNR authors in the period 1978 – 1998 were 48% CNR-owned (Guatta, 2002). This is, however, on average just one EPO patent more per year per university. Therefore, considering that the same can be seen in other countries, although on a much smaller scale ( (Mayer–Krahmer and Smoch, 1998, Tijssen, 2001; Washburn, 2005), the gap with the G7 countries remains considerable. In 2002, in Italy, patents with university ownership were 4 for each thousand researchers, in the United Kingdom 22, whilst in the U.S.A. already in 1999 they exceeded 40 (Abramo, 2007).

Although university research, compared to industrial research, in the spectrum of basic-applied research swings more towards basic, it is often conceived to produce new practical and technological knowledge of real interest to industry (Rosenberg and Nelson, 1994). It is legitimate to expect, therefore, that excellence in research gives rise to both a high intensity of publications (publications per researcher) and an equally high intensity of patenting.

Whilst studies on research productivity in American universities reveal that there exists a strong correlation between publications and patents, that is the universities with the greatest intensity of publications are also those with the greatest intensity in patenting (Adams and Griliches, 1998; Lach and Shankerman, 2003)[9] a similar research into the Italian reality reveals an absence of any form of correlation (Abramo, 2007). On the other hand, the vertiginous increase in patent production in American universities in these last twenty years, following the introduction of the Baye-Dole Act, cannot be entirely explained as an effect of increased research productivity; rather it is more likely a result of a more appropriate use of the forms of codification of new knowledge of interest to industry (more patents, fewer publications). It has been demonstrated that the increased intensity of patents did not happen following a shifting of gravity in research from basic to applied, not from an evolution of the portfolio of research projects towards fields of investigation with a higher patent fertility (Mowery et al., 2001; Thursby and Thursby, 2002).

The deductions to be made from the empirical evidence reported enables one to conclude that the productive efficiency of the Italian public research system is certainly not inferior to that of the other major industrialized countries; nevertheless, the recourse to scientific publication rather than to patent, as codification of new knowledge produced, is greatly superior to that of others.

The evidence that scientific literature is a source of information in the innovation process more for the large enterprises than for the small ones (Cohen et al., 2002b) and the small presence of the first in the Italian industrial system, contrary to the situation in other countries, easily lets one know who is likely to benefit from new Italian knowledge codified in publications, rather than in patents.

Having analyzed the attitude towards patenting in the Italian public research system, let us now examine the intensity of technology transfer, taking into account that the empirical literary evidence available for Italy is extremely thin on the ground and regards essentially patenting activity. Evidence shows that the small amount of new knowledge codified in patent form is transferred to the productive system in a proportion decidedly inferior to that of other countries. In particular, between 1996 and 2001, the CNR transferred 19,5% (2,4% abroad) of patents filed during that period (Abramo, 2006), that is 2.2 transfers per year for every thousand researchers; universities transferred 13% (1.1% abroad) of patents filed in the period 1998-2002, that is 0.2 transfers for every thousand Fte[10] researchers (Abramo and Pugini, 2005). The comparison between the CNR and the MIT for the period 1999-2001 shows that the CNR, research expenses being equal, transferred to the productive system 6.85% of patents transferred by the MIT. In the five year-period 1999-2003, American universities transferred on average 60% of the patents filed in that same period; the same percentage can be seen for British universities in 2002 (Nubs, 2003), whilst in 2003 Canadian universities transferred more patents than those filed that year (Autm, 2004).

It is more difficult to evaluate the transfer of knowledge in the form of trade-secrets. Contracts for the granting of know-how are fairly rare for the CNR and universities and are in general associated with the granting of a patent license. Transfer of knowledge is obviously to be foreseen in the case of research projects financed or co-financed by the private sector, especially where the financing body assures itself of the exclusive property rights of the research already during contractual negotiations.

In Italy, however, the shares of R&D expenditure financed by the private sector represents 4.76% of higher education R&D expenditure and 2.26% of public research bodies (Airi, 2004) against 5.5% of American universities (Oecd, 2006). On the evidence based on the indicators of technology transfer there can be added a direct indication deriving from an investigation by the Italian Institute of Statistics, Istat (2003). According to this organization, the Italian companies that introduced innovations in products or processes in the period 1998 – 2000 relegate public research bodies and universities to the last positions amongst the 10 possible sources of information constituting the basis for process innovation.

Last but not least, public research spin off rates, investigated by Abramo and Pugini, 2005 reveal too substantial delays, when compared to other countries. In the five-year period 1999 – 2003 each Italian university, on average, gave life to 0.03 spin offs[11] per year, against 2.6 on average for American universities. British universities in 2003 gave life to about 1 spin off per university.

3. Research methodology

Having presented the level of public to private technology transfer in Italy, we pass now to test the core hypothesis of our study, i.e. whether a mismatch exists between the private demand and the public supply of "new knowledge". To the authors' knowledge no national scale investigations of this kind have been reported in the literature.

A survey was conducted through a questionnaire which was sent to researchers employed in public research laboratories operating in different high-tech sectors. The relevant literature and the guidelines included in the Italian National Research Plan led us to select the following areas out of all those classified as high-tech: i) biotechnologies, pharmacy and biomedical technologies; ii) new materials; iii) nano-technologies, micro- and opto-electronics; and iv) robotics, sensor sciences and artificial intelligence. The choice of high-tech sectors rather then middle- or low-tech sectors is conservative: a possible mismatch is more likely to be found there, given the low-tech specialization of Italian industry. The reference source from which we extracted data for our study was the Science Citation Index® (SCI) database of Thomson Scientific, in Philadelphia. The database covers scientific publications in qualified international journals on the basis of the refereeing process and the screening of papers submitted for publishing[12]. Globally, SCI covers nearly six thousand international journals[13], 1772 of which are included in the four selected areas (Table 1).

*Table 1: Number of SCI journals in selected areas*

| Area | SCI journals |
|---|---|
| A - Biotechnologies, pharmacy and biomedical technologies | 729 |
| B - New materials | 576 |
| C - Nano-technologies, micro- and opto-electronics | 320 |
| D - Robotics, sensor sciences and artificial intelligence | 147 |
| Tot. | 1 772 |

The 1 772 total journals, distributed by area, have been ranked from the highest to the lowest *impact factor*, so as to identify the best 20 for each area. Considering the importance of the citations they received, those journals can be classified among the top 20 in terms of quality of published material. Table 2 shows the number of scientific publications by Italian authors in the period 2000-2002, by area and year, both as a total and as restricted to the top-20 journals.

*Table 2: Number of publications by Italian authors per area and per year (processing of SCI data)*

| Area | 2000 Total | 2000 top-20 only | 2001 Total | 2001 top-20 only | 2002 Total | 2002 top-20 only | 2000-2002 Total | 2000-2002 top-20 only | Selected |
|---|---|---|---|---|---|---|---|---|---|
| A | 4 191 | 284 | 4 329 | 387 | 4 097 | 173 | 12.617 | 844 | 319 |
| B | 1 763 | 128 | 2 132 | 175 | 1 827 | 202 | 5.722 | 505 | 191 |
| C | 2 017 | 316 | 2 487 | 365 | 2 301 | 372 | 6.805 | 1.053 | 398 |
| D | 201 | 70 | 227 | 84 | 220 | 87 | 648 | 241 | 92 |
| Total | 8 172 | 798 | 9 175 | 1 011 | 8 445 | 834 | 25 792 | 2 643 | 1 000 |
| Total production | 31 677 | | 33 685 | | 32 323 | | 97 685 | | |

On the whole, from 2000 to 2002, Italian researchers published 97 685 scientific papers in journals covered by the SCI (Tuzi, 2005). In particular, 25 792 (26.4%) belonged to the areas selected, with a yearly average of 4.85 papers per journal. Publications by Italian researchers in top-20 journals, in the same four areas, amounted to 2 643, with a yearly average of 11.01 papers per journal. Incidentally, the higher rate of publication intensity in top-20 journals by Italian authors with respect to the total number of journals is yet another indicator of the excellence of Italian public research. Through random sampling, 1 000 publications were selected out of the 2 643 in top-20 journals chosen for coverage: attention was paid to area representativeness. Table 3 sums up all the phases of the sampling process, and the data related to it.

Our questionnaire was then sent by e-mail to the first author of each of the 1 000 sampled articles. It comprised four sections related to information on respondent (*i*) and on his main research (*ii*); organization of research (*iii*) and transferability of findings (*iv*). For details see the methodological annex.

The global response rate was 44.7%. The pool had to be further reduced during post-codification, when irrelevant cases were excluded from examination. In particular, 42 questionnaires were discarded because they were filled in by researchers operating in different areas than those forming our frame of reference. It should be noted that the 405 final respondents represent over 1% of the population of Italian public researchers, representing a reliable sample[14].

*Table 3: Reference universe and research sample*

| | | |
|---|---|---|
| Overall publications (2000-2002) | 97 685 | 100% |
| Number of publications in selected areas | 25 792 | 26,4% |
| Number of publications in selected areas, top-20 journals only | 2 643 | 10,25% |
| Extracted sample (questionnaires sent) | 1 000 | 37,84% |
| Respondents | 447 | 44,7% |
| Valid responses | 405 | 90,6% |

## 4. Results of the analysis

This section will deal with the analysis of collected data. In order to verify the hypothesis that prompted our empirical research, we will first present the results of the analysis of data regarding the typological characterization of projects under consideration, in terms of type of research, affiliation of project team members and source of funding. We will then present our results regarding the expected modes of codification of the project findings, and illustrate data supporting our research hypothesis, with particular reference to industrial applicability of the project findings, the nationality of potential users and the existence of possible correlations with project typology. We conducted bivariate analyses by means of contingency tables and concentration indices. Chi-square tests were carried out in order to verify dependence among variables and statistical significance of the concentration indices.

### 4.1 Characterization of research projects

Data elaborations indicate that, according to respondents, the high-tech projects they were involved in concerned basic research in 48.6% of cases, and applied research in 45.4% of cases. Research on "generic technologies" [15] (under 5%) was much less common; the predominant tendency was towards frontier research (85.9%) rather than catch-up research (14.1%). As regards the latter, applied research is more frequent than basic research. A cross-check of sector data shows that the new materials area has a particularly high concentration of basic research and research directed towards development of new generic technologies. Applied research tends to concentrate in the remaining areas (biotechnologies, nano-technologies, robotics). As far as organization is concerned, in most cases (290, i.e. 71.6% of total) research teams turn out to comprise researchers affiliated to different institutions. Public research is thus strictly intramural only in 28.4% of cases. Data also show the central role of co-participation of private-company human resources in public research (12.9% of overall projects; 29.7% of cross-organizational projects). In 18.3% of cases, research teams include researchers affiliated to two separate organizations; in 26.4% of cases, the team extends to three organizations, and in 26.9% of cases, it includes four or more organizations at the same time (Figure 1). On average, every research project involves 2.65 research organizations.

The above indicates that the organizational nature of research teams is often extremely articulated and complex. As far as funding sources are concerned, only 11.1% of the projects do not resort to external sources, as against 59.3% of those that can count on additional funds from exclusively public sources, 7% obtaining support from private companies only, and nearly 23% of the projects receiving funding from both public and private sources.

*Figure 1: Frequency distribution of the number of organizations involved in joint-research projects*

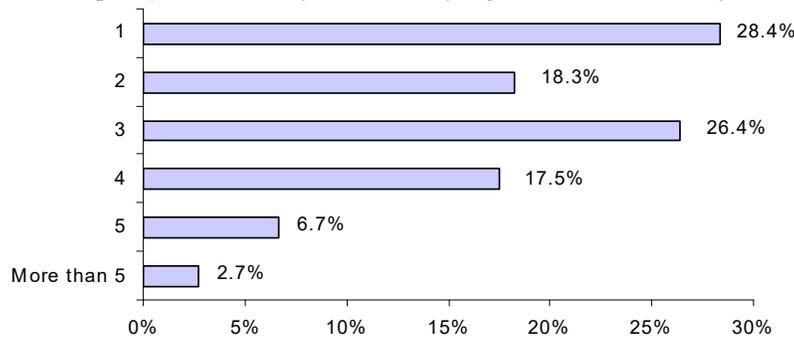

Table 4 shows the sources of external funding in relation to the type of research financed. Each cell reports the number of projects and their concentration: the evidence shows that basic research is mainly supported with additional public funds (concentration index = 113.5[16]), or can rely only on its on own resources, whereas private funding tends to concentrate on applied research projects, even when the projects attract public co-funding.

Table 4: Contingency table with absolute values and concentration indices (in brackets) per type of research and funding ($\chi^2 = 24.314$, p-value < 0.0005)

| ExternalFunding | Type of research | | | | Total |
|---|---|---|---|---|---|
| | Basic | Applied | Generic technologies | Other | |
| None | 24 (109.6) | 18 (88.0) | 3 (135.0) | 0 (0.0) | 45 (100.0) |
| Public | 132 (113.5) | 92 (84.7) | 12 (101.7) | 3 (127.1) | 239 (100.0) |
| Private | 5 (36.7) | 23 (180.8) | 0 (0.0) | 0 (0.0) | 28 (100.0) |
| Public and private | 36 (79.6) | 51 (120.7) | 5 (108.9) | 1 (108.9) | 93 (100.0) |
| Total | 197 (100.0) | 184 (100.0) | 20 (100.0) | 4 (100.0) | 405 (100.0) |

The frequency distribution of the number of public funding sources (Figure 2) shows that research projects receive additional public funding from two or more different institutions in one out of three cases.

*Figure 2: Frequency distribution of the number of public funding sources per single research project*

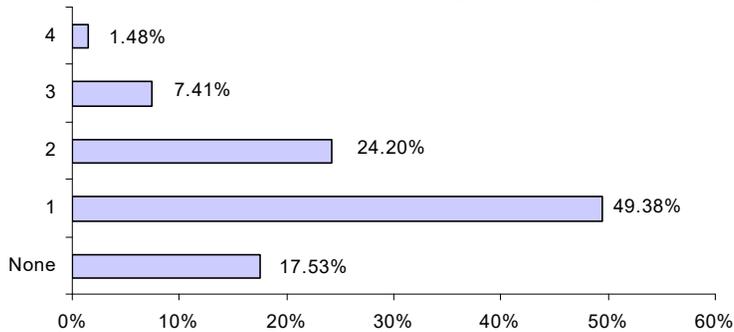

The analysis of private funding indicates that the area where that is most frequent is biotechnologies, followed by robotics (Table 5). New materials rank last; as, we have seen, they are marked by a predominance of basic research. Due to the type of research, entailing long-term planning of possible industrial and multi-sector commercial applications, this discipline seems to attract direct private funding less readily than others.

*Table 5: Private funding allocation*

| Area | Number of research projects funded by private organizations | % |
|---|---|---|
| A | 51 (of 144) | 35.4 |
| B | 23 (of 99) | 23.2 |
| C | 30 (of 116) | 25.9 |
| D | 16 (of 46) | 34.8 |
| Total | 120 (of 405) | 100 |

Domestic companies (large and small alike) are the ones that give the largest share of private support to public research projects, but there is also a significant presence of foreign private organizations, sponsoring 30% of private-funded projects (Table 6).

*Table 6: Distribution of type of private organizations funding public research projects*

| Private organization type | Project funded | As total of project funded by private (120) | As total of private organizations funding (151) |
|---|---|---|---|
| National large company | 38 | 31.67% | 25.17% |
| National small-medium enterprise | 35 | 29.17% | 23.18% |
| Telethon + ONLUS | 27 | 22.50% | 17.88% |
| Foreign large company | 25 | 20.83% | 16.56% |
| Bank or foundations | 8 | 6.67% | 5.30% |
| International small-medium enterprise | 7 | 5.83% | 4.64% |
| National enterprise association | 7 | 5.83% | 4.64% |
| International enterprise association | 4 | 3.33% | 2.65% |
| Totale | 151 | | 100% |

Among national sponsors, associations and foundations set up in recent years to support research on a number of particularly serious diseases, take a central role in the picture of private funding, and they account for up to 23% of the total number of privately supported projects[17]. The last standing in the rank is occupied by banks and banking foundations. Unlike public funding, private funding is more frequently associated to exclusiveness. 25 projects only (out of total 120) receive private funds from more than one source at the same time (Figure 3).

*Figure 3: Distribution of the number of private funding sources per research project*

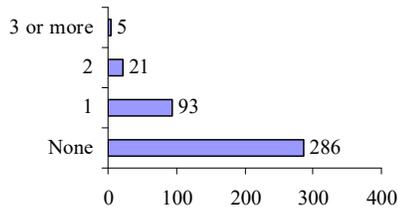

### 4.2 Modes of codification of research results

Before analyzing the distribution of answers regarding transferability of project findings to the industrial system and trying to find an answer to the research question our study is based on, we will present the results of the analysis of the modes of codification of project outputs. We will subsequently be able to carry out cross-analyses and deal with particular points in more depth. Table 7 shows that the preferred form of codification of research findings is the scientific publication (81% of responses).

*Table 7: Distribution of modes of codification of research results*

| Output codification | Primary | Secondary |
|---|---|---|
| Scientific article | 328 (80.99%) | 63 (15.56%) |
| Prototype | 50 (12.35%) | 74 (18.27%) |
| Patent | 21 (5.19%) | 176 (43.46%) |
| Trade secret | 4 (0.99%) | 21 (5.19%) |
| Other | 2 (0.49%) | 30 (7.41%) |
| Missing | - | 41 (10.12%) |
| Total | 405 (100%) | 405 (100%) |

In 12% of cases, the research will lead to the development of a demonstrative prototype, whereas only in 5% of cases is patent the preferred form of codification of project results. The comparative analysis of occurrences of the primary and secondary forms is rather interesting[18]. The fourth column of Table 7 indicates that patents rank first as secondary form of codification, followed by prototypes. At any rate, if the sole responses on primary forms are considered, the expected patent intensity (number of patents per 1,000 publications) of the research projects under consideration amounts to 64. This figure is higher by one order of magnitude than that indicated by the Osservatorio sulla Ricerca Pubblica (Observatory on Italian Public Research)[19], which, for the period 2000-2002, covered the output of the national public research system in terms of publications (around 80,000) and patents filed by Italian public institutions (477). In other words, the actual number of filed patents is below the number that one would expect according to the interviewees' responses. That difference may have three different reasons:
- the areas under consideration (high tech) are more "fertile" than the average, in

terms of patentability of research results;
- inventors tend to file patents directly and claim ownership, as current laws entitle them to the right of ownership of the findings of research conducted within their institutions;
- ex-ante intentions and ex-post decisions diverge: various research outputs may well be codified and protected through patents but, for some reason, are not.

There is no empirical evidence confirming or denying the first motivation. It is certainly conceivable that a more frequent recourse to publication depends, as we will see shortly, on the typology of research (basic vs. applied) rather than on the "technological level" of a particular area. The second hypothesis relies on a study by Balconi et al. (2004), which concluded that, at university level, patent applications authored by university personnel are filed by third parties substantially more often than by the universities themselves. The third motivation would seem to indicate that a link in the chain of the process of public-to-private technology transfer is structurally weak; that this weakness in the system would lead public researchers to favour scientific publications as a tool to codify and disseminate the new knowledge developed in their research. The concentration analysis also points to specific differences within single research areas (Table 8)[20].

Scientific publication is the form most frequently used by researchers to codify and disseminate a research output. This occurs most often in the area of biotechnologies, whereas the patent tends to be the most preferred form of codification of the outputs of projects regarding new materials. The area of nano-technologies is marked by a certain concentration of preferences for prototypes; the latter is more often used in robotics than in the remaining areas. The modes of codification in alternative to publication occur more frequently in applied research, as the data in Table 9 show. Projects enjoying only private funding see a high concentration of patents, prototypes and industrial secrets as forms of output codifications (Table 10). A good 16 out of the 28 projects under consideration (57%) have a proprietary form of codification of their results.

*Table 8: Contingency table with absolute values and concentration indices (in brackets) per output primary mode of codification and research area($\chi^2$ = 22.534, p-value < 0.032)*

| Output primary mode of codification | Research area | | | | Total |
|---|---|---|---|---|---|
| | Bio-tech | New materials | Nano-tech | Robotics | |
| Scientific article | 122 (104.6) | 83 (103.5) | 90 (95.8) | 33 (88.6) | 328 (100.0) |
| Patent | 8 (107.1) | 7 (136.4) | 4 (66.5) | 2 (83.9) | 21 (100.0) |
| Prototype | 12 (67.5) | 7 (57.3) | 20 (139.7) | 11 (193.7) | 50 (100.0) |
| Trade secret* | 1 (70.3) | 1 (102.3) | 2 (174.6) | 0 (0.0) | 4 (100.0) |
| Other | 1 (140.6) | 1 (204.5) | 0 (0.0) | 0 (0.0) | 2 (100.0) |
| Total | 144 (100.0) | 99 (100.0) | 116 (100.0) | 46 (100.0) | 405 (100.0) |

* *not considered in the test of independence*

*Table 9: Contingency table with absolute values and concentration indices (in brackets) per output primary mode of codification and research type($\chi^2 = 59.486$, p-value $< 5.8E-11$)*

| Output primary mode of codification | Research type | | | | Total |
|---|---|---|---|---|---|
| | Basic | Applied | Generic technologies | Other | |
| Scientific article | 186 (116.6) | 119 (79.9) | 19 (117.3) | 4 (123.5) | 328 (100.0) |
| Patent | 3 (29.4) | 17 (178.2) | 1 (96.4) | 0 (0.0) | 21 (100.0) |
| Prototype | 6 (24.7) | 44 (193.7) | 0 (0.0) | 0 (0.0) | 50 (100.0) |
| Trade secret | 1 (51.4) | 3 (165.1) | 0 (0.0) | 0 (0.0) | 4 (100.0) |
| Other | 1 (102.8) | 1 (110.1) | 0 (0.0) | 0 (0.0) | 2 (100.0) |
| Total | 197 (100.0) | 184 (100.0) | 20 (100.0) | 4 (100.0) | 405 (100.0) |

*Table 10: Contingency table with absolute values and concentration indices (in brackets) per output primary mode of codification and funding source ($\chi^2 = 51.880$, p-value $< 6.6E-7$)*

| Output primary mode of codification | Funding source | | | | Total |
|---|---|---|---|---|---|
| | Public | Private | Publ.+priv. | None | |
| Scientific article | 207 (106.9) | 12 (52.9) | 71 (94.3) | 38 (104.3) | 328 (100.0) |
| Patent | 9 (72.6) | 6 (413.3) | 3 (62.2) | 3 (128.6) | 21 (100.0) |
| Prototype | 21 (71.2) | 8 (231.4) | 17 (148.1) | 4 (72.0) | 50 (100.0) |
| Trade secret | 1 (42.4) | 2 (723.2) | 1 (108.9) | 0 (0.0) | 4 (100.0) |
| Other | 1 (84.7) | 0 (0.0) | 1 (217.7) | 0 (0.0) | 2 (100.0) |
| Total | 239 (100.0) | 28 (100.0) | 93 (100.0) | 45 (100.0) | 405 (100.0) |

### 4.3 Transferability of findings from research projects

Let us now come to the main objective of our study, i.e. verifying the matching between the supply of research results issuing from the activities of public institutions and the demand from private users. We find that, as regards industrial applications, about 72% of the interviewees (291 out of 405) stated that findings from their research was of immediate industrial interest; as expected, the applicability is more frequent in applied research projects than in other typologies (Table 11). In one out of three cases, the potential user is a foreign company, usually European (Table 12).

*Table 11: Contingency table with absolute values and concentration indices i (in brackets) per research type and industrial applicability ($\chi^2 = 37.866$, p-value $< 6E-9$)*

| Industrial applicability | Research type | | | | Total |
|---|---|---|---|---|---|
| | Basic | Applied | Generic technologies | Other | |
| YES | 117 (82.7) | 161 (121.8) | 13 (90.5) | 0 (0.0) | 291 (100.0) |
| NO | 80 (144.3) | 23 (44.4) | 7 (124.3) | 4 (355.3) | 114 (100.0) |
| Total | 197 (100.0) | 184 (100.0) | 20 (100.0) | 4 (100.0) | 405 (100.0) |

*Table 12: Distribution of industrial potential users of research findings*

| Nationality of potential industrial user | Cases | % |
|---|---|---|
| Italian | 195 | 67.01 |
| European (EU-15) | 80 | 27.49 |
| USA | 11 | 3.78 |
| Other | 5 | 1.72 |
| Total | 291 | 100 |

Private-funded research has industrial applications especially in the country where the funding comes from (Table 13).

Table 13: Contingency table with absolute values and concentration indices (in brackets) per nationality of potential industrial users and of private funding organizations ($\chi^2 = 46.769$, p-value < 3.9E-10)

| Nationality of potential industrial users | Nationality of private funding organizations | | | | Total |
|---|---|---|---|---|---|
| | None | Italy | Foreign | Italy +Foreign | |
| Italian | 124 (95.9) | 55 (120.7) | 7 (65.3) | 9 (95.9) | 195 (100.0) |
| Foreign | 69 (108.4) | 13 (58.0) | 9 (170.5) | 5 (108.3) | 96 (100.0) |
| Total | 193 (100.0) | 68 (100.0) | 16 (100.0) | 14 (100.0) | 291 (100.0) |

In those projects enjoying private national and foreign co-funding, the industrial applications are envisaged in foreign countries in 5 out of 14 cases. Besides, projects that obtain no private funding (while using national public resources) are shown to have findings exploitable abroad in 69 out of 193 cases. The data in Table 14 indicate that the projects the findings of which are preferably codified and disseminated through scientific publications find their potential industrial users in foreign organizations in 77 cases out of 216 (36% of the total).

Table 14: Contingency table with absolute values and concentration indices (in brackets) per nationality of potential industrial users and output primary mode of codification ($\chi^2 = 8.086$, p-value < 0.232)

| Output primary mode of codification | Nationality of potential industrial users | | | | Total |
|---|---|---|---|---|---|
| | ITA | EU | USA | Other | |
| Scientific article | 139 (96.0) | 64 (107.8) | 11 (134.7) | 2 (53.9) | 216 (100.0) |
| Patent | 16 (113.7) | 5 (86.6) | 0 (0.0) | 0 (0.0) | 21 (100.0) |
| Prototype | 36 (109.6) | 11 (81.7) | 0 (0.0) | 2 (237.6) | 49 (100.0) |
| Trade secret* | 2 (99.5) | 0 (0.0) | 0 (0.0) | 1 (1940) | 3 (100.0) |
| Other | 2 (149.2) | 0 (0.0) | 0 (0.0) | 0 (0.0) | 2 (100.0) |
| Total | 195 (100.0) | 80 (100.0) | 11 (100.0) | 5 (100.0) | 291 (100.0) |

*not considered in the test of independence*

Finally, a cross-analysis of the data on the nationality of potential users and research typology indicates that research on generic technologies, while not numerous, usually find industrial applications abroad (Table 15).

Table 15: Contingency table with absolute values and concentration indices (in brackets) per research type and nationality of potential industrial user ($\chi^2 = 36.416$, p-value < 1.3E-8)

| Nationality of potential industrial user | Research type | | | Total |
|---|---|---|---|---|
| | Basic | Applied | Generic technologies | |
| Italian | 77 (98.2) | 112 (103.8) | 6 (68.9) | 195 (100.0) |
| Foreign | 40 (103.6) | 49 (92.3) | 7 (163.2) | 96 (100.0) |
| Total | 117 (100.0) | 161 (100.0) | 13 (100.0) | 291 (100.0) |

Further cross-analyses on the sub-sample, including those research projects with possible industrial applications, which, according to respondents, are of interest to foreign users, did not result in more significant conclusions. In terms of output codification, scientific publications appear to guarantee the best system of dissemination and valorization of research findings, especially when possible immediate industrial applications are lacking (Table 16).

Data show that the use of patents and prototypes as modes of codification is mainly concentrated on projects that result in research outputs interesting for their industrial applicability. The data regarding the sub-sample comprising the 291 projects with findings of an immediate industrial application indicate that the most frequent primary form of codification remains the scientific publication (74% of cases). Patents, instead, occur in 7.2% of all cases as primary mode, and in 45.4% as secondary mode.

*Table 16: Contingency table with absolute values and concentration indices (in brackets) per output primary mode of codification and industrial applicability($\chi^2$ = 33.705, p-value < 8.6E-7)*

| Output primary mode of codification | Industrial applicability | | Total |
|---|---|---|---|
| | YES | NO | |
| Scientific article | 216 (91.7) | 112 (121.3) | 328 (100.0) |
| Patent | 21 (139.2) | 0 (0.0) | 21 (100.0) |
| Prototype | 49 (136.4) | 1 (7.1) | 50 (100.0) |
| Trade secret | 3 (104.4) | 1 (88.8) | 4 (100.0) |
| Other | 2 (139.2) | 0 (0.0) | 2 (100.0) |
| Total | 291 (100.0) | 114 (100.0) | 405 (100.0) |

## 5. Discussion and conclusions

This study has shown a number of peculiarities in the structure of Italian public research, at least in the high-tech areas selected. First of all, the higher concentration of publications in high-impact-factor journals is another clear indicator of the level of excellence of research conducted in Italy at public level. This confirms Italy's striking performance in scientific productivity among G7 countries.

The weight of basic research, 48.6% (and its predominance over applied research, which is at 45.4%) appears to be high if compared to national averages found in literature, although the public character of the research institutions and the particular areas selected partly account for that distribution, the dividing line of which is still difficult to identify. Furthermore, a high level of cooperation among research groups affiliated to different institutions was observed. Projects involving research groups within a single institution are the exception rather than the rule.

The organization of projects is often extremely articulated and complex, and sometimes involves four or five different research institutions at the same time. The reasons for this are that synergies are also sought outside, and that recourse to external funding on a competitive basis is increasingly frequent (90% of the projects is granted such funds): partnerships, often a prerequisite to have access to external funding (for example the EU Framework Programmes), enhance the representative strength of the proposing institution and the chances of accessing funds. It has also been noticed that nearly 30% of the projects receive funds from private sources, of which 7% in exclusive form, and 23% along with public funding. The private industrial system, however, accounts for only 4,76% of the total revenues of Italian universities and 2.26% for public research institutions (Airi 2004), as against 5.0% of American universities (Oecd, 2006). It would thus seem that funding tends to concentrate on projects of excellence (our sample comprises authors of publications in top-20 journals), covering only a marginal part of the total costs of a research project. On the other hand, the aforementioned study by Balconi et al. (2004) shows that, at university level, patents owned by third parties are filed substantially more often than patents owned by the universities to which the inventors are affiliated (77% are owned by companies, 16% of which abroad). This goes to prove that the value transferred to those who have most probably co-financed specific research is not in proportion to their total expenses. It is also evident that Italian public researchers consider scientific publications (81% of cases) the best (or easiest) way to codify new knowledge developed throughout their research and to disseminate and exploit it.

The patent is certainly a tool frequently used as a form of codification of findings, but only as a second-best option. This leads to two considerations. First of all, it may well be the case that Italian researchers are not aware of (all or part of) the practices and procedures regarding patent protection[21]. Such practices and procedures include an obligation not to disclose results publicly before filing a patent application, a violation of which would invalidate the patent grant. Secondly, the prevalence of publications over patents as the primary form of codification of research findings poses a problem of externality (at least for the 74% of cases, in which the authors believe that the results of a research project will have immediate industrial applicability).

Experts in economic geography agree in saying that geographical "proximity" between companies and research institutions may function as a catalyst in the processes of creation and dissemination of new knowledge, and therefore in the development and economic growth of a particular region. The intensity of the "proximity" effect, however, is deeply affected by the form of codification that the exchanged knowledge assumes; in particular, due to its nature of public good, a scientific paper, especially one published in an international journal, generates non-localized spill-overs. Furthermore, as has been shown (Cohen et al., 2002b), larger companies use that source to feed their internal innovation processes more often than smaller ones, and Italy's industrial system is admittedly marked by the extensive presence of small and medium companies; it is thus plain to see that publication favours the technological competitiveness of foreign rather than Italian companies. Paradoxically, the indisputable scientific excellence of Italian researchers, if not exploited properly, might have a boomerang effect in the long run, thwarting instead of supporting the technological competitiveness of the Italian industrial system with respect to that of other countries.

As regards potential spillovers on the industrial system, around 72% of the ongoing projects foresee immediate industrial applicability. This means that part of basic research leads to results useful industry. This is especially plausible when scientific research is of the gap-filling sort, that is when technological progress overtakes scientific progress, and further technological advances need deeper understanding of underlying phenomena, leading to what is known as gap-filling scientific research (Allen, 1984), the findings of which are of immediate interest for (corporate) technologists. A good third of the projects with immediate industrial applications has no Italian partners able to exploit their results.

The implication is that economic and strategic benefits of about 24% of high-tech research employing national public funding will be enjoyed mainly by the economies of other countries, unless a relevant number of research-based spin-offs or start-ups are created. There are more relevant factors: i) researchers tend to favour publications over patents, even for the findings of industrial interest, for which the latter form of codification would be more suitable; ii) the percentage of patents transferred abroad (Abramo, 2006) on the total is not negligible (12% for the CNR and 9% for universities); and iii) the number of public-research-based spin-offs in Italy is extremely low. All this means that a really small portion of the economic and social returns on public research is appropriated by Italy.

Such a result unequivocally denotes a misalignment between research policies and industrial policies in Italy. Scientific excellence is of little use, in the absence of an industrial system able to harvest its benefits. This gives rise to the need to formulate the priorities of national research (and the relevant interventions and incentive systems), based, on the one hand, on the ability of Italian industries to exploit the findings resulting from that research, and, on the other hand, on strategies of sectorial development. Whereby a large portion of the national science base is not matched by industry demand, if a national competitive presence is to be established or reinforced, industrial entrepreneurship (research-based start-ups) and academic entrepreneurship (spin-offs) are to be fostered, in order to avoid a "research results drain", following the (more renowned) "brain drain".

Finally, for two thirds of the research projects the findings of which have immediate industrial applicability, a potential Italian industrial partner is found, according to interviewees. Due to the high specialization of Italian low-tech industry, it is expected that within public research projects as a whole (high, medium and low-tech), the share of those affecting Italian industry is even higher. These (conservative) data dispel the idea that there is a significant mismatch between public research supply and private domestic demand, which may justify the scarce technology transfer in Italy.

In a context where: *i* for 24% of all public research projects no national company is found that can immediately exploit the relevant results; *ii* in 48% of the cases in which a partner is found, the research results are mostly codified as publications; *and iii* only a small share of the few patents that are filed are licensed, while the current debate focuses on the limited resources allocated for public research, the authors urge policy makers to shift their attention from "how loud and clear", to "for whom" the bell tolls. The scientific excellence of a country is a fleeting one if it does not translate into development and better quality of life for the tax-payers who nurture it.

**Notes**

[1] The chronic and increasing deficit of the Italian technological balance of payments for patents, passed from 270 million of euro in 1998 to 500 million euro in 2003.

[2] The inadequacy of the "transfer system" at the supply end, which is the third possible cause of low technology transfer, along with scarce domestic demand and demand-supply mismatch, was studied in Abramo (1998 and 2006).

[3] Gallino describes how flourishing high-tech sectors of the Italian industry, such as the chemical, computer, airplane, electronic industries either disappeared or irremediably shrunk..

[4] . Different proportions of public research scientists in the various countries have been accounted for. A detailed description of the calculation methodology and assumptions made can be found in Abramo, 2006.

[5] Incidentally, it is worth mentioning that the evidence provided by Bonaccorsi (2007) of the weak performance of European science in the upper tail of scientific quality, cannot be ascribed to Italy.

[6] The Third European Report on Science and Technology Indicators (European Commission, 2003) shows that the distribution of publications in Italy in the various scientific journals is similar to that of the other G7 countries.

[7] It is a widespread opinion that the systems of recruitment and career progression in the Italian public

research institutions are less meritocratic than those in other G7 countries, and this inevitably has a negative impact on the quality and general organization of the work force, and the brain drain. Direct empirical evidence is supplied, at least for the sector of economic sciences, by Gagliarducci et al., 2005. Another indicator is the number of Italian Nobel prize winners in scientific areas; one has to go back to 1963 to find an Italian scientist, Giulio Natta, a receiver of the Nobel prize for studies conducted in Italy.

[8] 2001 has been chosen as a year of reference because it was starting from that year that the new Italian legislation on patents assigned the intellectual property rights of research results to the public researcher, rather than to the employer.

[9] The evidence has been confirmed by a very recent survey relating to the Catholic University of Leuven (Van Looy et al., 2006).

[10] Fte, full time equivalent.

[11] For comparative reasons the Autm definition of an academic spin off was adopted: a new enterprise set up following the acquisition of a university patent, independently of the participation of university personnel or of the university in the company capital.

[12] This qualification is a guarantee on the quality of a journal.

[13] Thomson Scientific data bases are universally acknowledged as the chief reference in bibliometric research.

[14] A sample of 400 units ensures, with a probability of 95.46%, results extendable to the universe with an error margin not exceeding ± 5%, provided the distribution of responses is normal. We made a test run, which showed a meaningful representativeness of the sample with respect to the four investigated research areas (p-value 0,95).

[15] The questionnaire included a glossary with the technical terms found in it. In particular, *basic research* is defined as experimental or theoretical work undertaken primarily to acquire new knowledge of the underlying foundations of phenomena and observable facts, without any particular application or use in sight *Applied research* is also original investigation undertaken in order to acquire new knowledge. It is, however, directed primarily towards a specific practical aim or objective. "Pre-industrial research", also called "research in *generic technologies*", falls between basic and applied research, and is defined as a type of research envisaging potential market application, though not yet focusing on a specific area.

[16] $113.5 = \dfrac{132/197}{239/405} \times 100$ Because the concentration index is above 100, the trait "public funding" tends to concentrate on the type of project "basic research projects".

[17] That type of funding concentrates, for obvious reasons, on biotechnologies only.

[18] As regards forms of research output codification, the questionnaire included two blanks, one for the primary form and one for the secondary form.

[19] More detailed information on how the data base was set up at the research lab the authors belong to, may be found in Abramo et al., 2008.

[20] Concentration indices should not be misinterpreted: in all areas, scientific publications are the form of codification and dissemination of research outputs most frequently used by researchers. The data only point to the fact that, in few areas, such predominance is more evident than in others.

[21] It must be said that direct costs (including patent taxes and consultancy) and indirect costs (time devoted to writing documents, and publication delay) of patenting are substantially higher than those of publication.

## ANNEX

### The questionnaire
The questionnaire comprised four sections, with one-response-only closed questions.

A) Respondent
A.1) First name and last name
A.2) Affiliation

B) Information about current research
B.1) Project title
B.2) Field of research (bio-tech, new materials, nano-tech, robotics)
B.3) Tipology:
- Basic, applied, generic technology, other
- Catch-up vs frontier

C) Organization
C.1) Research team (intramural or extramural)
C.2) For extramural research: affiliation of other team members
C.3) Extra funding: public and/or private resources

D) Results
D.1) Primary and secondary form of codification (scientific article, patent, prototype, trade secret, other)
D.2) Industrial applicability
D.3) Nationality of potential industrial user

### Submission
The questionnaire was created in HTML-PHP format so as to enable automatic on-line response collection. The Java script support guarantees the interactivity of the questionnaire: the blank fields are generated progressively and visualized according to the answers entered in previous fields by each respondent.

An e-mail message was sent to the sampled researchers, explaining the objectives of the research and inviting them to fill in the questionnaire by simply following a link to a specially designed on-line form. An earlier version of this article was sent back to all respondents to elicit feedbacks and as a token of appreciation for their collaboration.

### Elaboration
The post-codification process aimed at identifying irrelevant cases and excluding them from examination. In particular, 42 questionnaires were discarded because they were filled in by researchers operating in areas different from those falling in our frame of reference. In fact, it is possible, although exceptional, that a researcher publishes findings of his research in a journal specializing in an area which is not directly related to his research interests. It is also occasionally possible that due to homonymy, the questionnaire was sent to researchers not included in the research target.

**Biographies of authors**

Giovanni Abramo holds a Laurea in Electrical Engineering from the University of Rome, Italy and a Master of Science in Management from the Massachusetts Institute of Technology, USA. He is senior research scientist at the National Research Council of Italy and member of the faculty of Engineering, Dept of Management, of the University of Rome, where he teaches Applied Economics in the undergraduate programme in Engineering, and Strategic Management in the master's programme in management. His interests in technology transfer and research evaluation led him to co-found the Laboratory for Studies of Research and Technology Transfer. He has served as a consultant to a number of international institutions and industry.

Ciriaco Andrea D'Angelo is assistant professor at the Dept of Management- School of Engineering of the University of Rome "Tor Vergata", Italy where he holds a PhD in Management in 1999. At the same University, he teaches Marketing courses in the undergraduate programme in Engineering and in the Master's programme in Management. He co-founded the Laboratory for Studies of Research and Technology

Transfer where he carries out research on industrial marketing, technology transfer, research evaluation, bibliometrics.